\documentclass[aps,twocolumn,amsmath,amssymb,reprint,numbers,superscriptaddress,noeprint]{revtex4-1}

\usepackage{makecell}
\usepackage{soul}
\usepackage{geometry}
\usepackage{graphicx}
\usepackage{booktabs}
\usepackage{array}
\usepackage{relsize}
\usepackage{color}
\usepackage{xcolor}
\usepackage{eqnarray}
\usepackage{bm}
\usepackage{textcase}
\usepackage{textcomp}
\usepackage{braket}
\usepackage{setspace}
\usepackage{mathtools}

\usepackage{hyperref}
\hypersetup{
colorlinks = true,
linkcolor = blue,
citecolor = blue,
urlcolor = blue
}

\definecolor{cola}{rgb}{0.7,0.1,0.1}
\definecolor{colb}{rgb}{0.6,0.6,0}
\definecolor{colc}{rgb}{0.3,0.7,0}
\definecolor{cold}{rgb}{0,0.35,0.75}
\definecolor{cole}{rgb}{0.63, 0.13, 0.94}
\definecolor{colf}{rgb}{0.0, 0.6, 0.0}

\newcommand{\xiayu}[1]{{\color{black} #1}}

\newcommand{\kater}[1]{{\color{black} #1}}

\def\be{\begin{equation}}
\def\ee{\end{equation}}

\newcommand{\dd}{\mathrm{d}}

\newcommand{\prt}[1]{\left(#1\right)}

\newcommand{\intmp}{\int_{-\infty}^{+\infty}}

\newcommand{\dg}{\dagger}

\newcommand{\prtq}[1]{\left[#1\right]}

\usepackage{physics}
\usepackage{amsfonts}
\usepackage{verbatim}
\usepackage{amsmath}
\usepackage{csquotes}
\usepackage{color}
\usepackage{soul}
\usepackage{amsthm}
\usepackage{bm}
\usepackage{graphicx}
\usepackage[normalem]{ulem}

\begin{document}

\title[]{Quantum energetics of a non-commuting measurement}

\author{Xiayu Linpeng}
\affiliation{International Quantum Academy, Futian District, Shenzhen 518048, China}
\affiliation{Department of Physics, Washington University, St. Louis, Missouri 63130, USA}
\author{Nicol\`o Piccione}
\affiliation{Universit\'e Grenoble Alpes, CNRS, Grenoble INP, Institut N\'eel, 38000 Grenoble, France}
\author{Maria Maffei}
\affiliation{Universit\'e Grenoble Alpes, CNRS, Grenoble INP, Institut N\'eel, 38000 Grenoble, France}
\author{L\'ea Bresque}
\affiliation{Universit\'e Grenoble Alpes, CNRS, Grenoble INP, Institut N\'eel, 38000 Grenoble, France}
\author{Samyak P. Prasad}
\affiliation{Universit\'e Grenoble Alpes, CNRS, Grenoble INP, Institut N\'eel, 38000 Grenoble, France}
\affiliation{MajuLab, CNRS-UCA-SU-NUS-NTU International Joint Research Laboratory}
\affiliation{Centre for Quantum Technologies, National University of Singapore, 117543 Singapore, Singapore}
\author{Andrew N. Jordan}
\affiliation{Institute for Quantum Studies, Chapman University, Orange, California 92866, USA}
\affiliation{Department of Physics and Astronomy, University of Rochester, Rochester, New York 14627, USA}
\author{Alexia Auff\`eves}
\affiliation{Universit\'e Grenoble Alpes, CNRS, Grenoble INP, Institut N\'eel, 38000 Grenoble, France}
\affiliation{MajuLab, CNRS-UCA-SU-NUS-NTU International Joint Research Laboratory}
\affiliation{Centre for Quantum Technologies, National University of Singapore, 117543 Singapore, Singapore}

\author{Kater W. Murch}
\affiliation{Department of Physics, Washington University, St. Louis, Missouri 63130, USA}

\begin{abstract}
When a measurement observable does not commute with a quantum system's Hamiltonian, the energy of the measured system is typically not conserved during the measurement. Instead, energy can be transferred between the measured system and the meter. In this work, we experimentally investigate the energetics of non-commuting measurements in a circuit quantum electrodynamics system containing a transmon qubit embedded in a 3D microwave cavity. We show through spectral analysis of the cavity photons that a frequency shift is imparted on the probe, in balance with the associated energy changes of the qubit. Our experiment provides new insights into foundations of quantum measurement, as well as a better understanding of the key mechanisms at play in quantum energetics.
\end{abstract}

\date{\today}

\maketitle

\section{Introduction}

The incompatibility of different observables in quantum mechanics is fundamental to its structure, and underlies its mysteries and limitations. These include wave-particle duality, uncertainty relations \cite{Robertson1929,Maassen1988,Monroe2021}, quantum measurement limits \cite{RevModPhys.82.1155}, non-locality, and even govern the controllability of quantum systems \cite{Huang1983}.
In particular, operator incompatibility plays a key role in the energetics of quantum measurement.
Indeed, quantum measurements are energy-preserving for the system under
measurement, provided that the measurement observable
commutes with the system Hamiltonian.
Conversely, the non-commuting case has been theoretically studied
extensively in the literature, with the Wigner, Araki,
and Yanase (WAY) theorem \cite{Araki1960,Yanase1961,ref:Ahmadi2013way} being one of the most
notable results. This theorem states that a perfect projective
measurement is not possible if the measurement
observable does not commute with additive conserved
quantities, such as the total energy of system and quantum
meter. In contrast, when there is an incompatibility
between the total energy operator and a measured
observable, anomalous energy changes in the measured system can occur, sometimes dubbed ``quantum heat'' or ``measurement energy'' \cite{Elouard2017, ref:Rogers2022pqe}.
The interplay between measurement incompatibility and energy forms the basis for new studies in quantum energetics where measurements can be used as a source of fuel for quantum measurement engines \cite{ref:Yanik2022tqm,ref:Aharonov2023clf,Elouard_maxwell, Elouard_Jordan,Yi2017,ding2018measurement,JEA,Mohammady2017, Bresque2021, Campisi, Buffoni,ref:Jussiau2023mbq,ref:Stevens2022esq}. However, with few experimental studies so-far investigating the detailed energy balance of such incompatible measurements, the fundamental mechanisms at play in such situations has remained elusive. In this paper, we investigate the energy balance of measurements that fail to commute with the system Hamiltonian, offering experimental evidence of the energy exchange between the measured quantum system and the meter. Our work provides new insights into the energetics of measurement and highlight connections between measurement fuel and paradigmatic dynamical effects in quantum measurement such as the quantum Zeno effect~\cite{Vijay2012,Slichter2016,Harrington2017}.

\section{Construction of a non-commuting measurement in circuit QED}

The recent advances in quantum science with the superconducting circuit quantum electrodynamics (circuit QED) architecture \cite{ref:Blais2004cqe, ref:Blais2021cqe} have been enabled by the ability to perform high fidelity non-demolition measurements, despite the low energy scale of the quantum bits. By and large, these measurements rely on the dispersive interaction between the qubit and a microwave resonator \cite{Boissonneault2009}, where the interaction is approximately given by $H_{\mathrm{int}}=\chi a^{\dagger} a \sigma_z$, where $\chi$ is the dispersive shift, $a^\dagger a$ is the resonator photon number operator, $\sigma_z$ is the Pauli operator that acts on the qubit in the energy basis, and we set $\hbar = 1$ in all equations for simplicity. This dispersive interaction shifts the phase of the resonator photons depending on $\langle \sigma_z \rangle$, thus providing a measurement observable in the $Z$ basis. As this interaction commutes with qubit energy operator $H_{\mathrm{q}} = \omega \sigma_z/2$, where $\omega$ is the qubit frequency, this interaction provides a system-energy-preserving observable for quantum non-demolition measurements.


\begin{figure}[!t]
  \centering
  \includegraphics[width=0.5\textwidth]{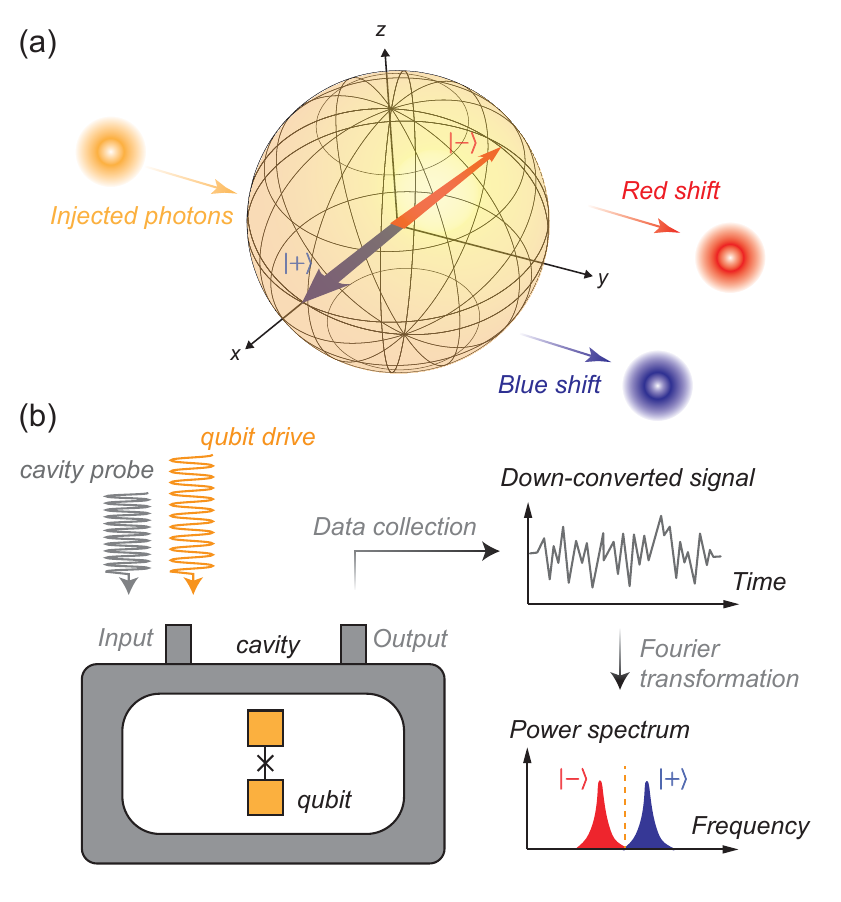}
  \caption{\label{fig:setup} (a) Illustration of the energy shift for the measurement photons after a non-commuting measurement. If the qubit is initially in the $\ket{+}$ state, the injected photon undergoes a blue shift, while for the $\ket{-}$ state, it undergoes a red shift. (b) Schematic of the experimental setup; a continuous qubit drive generates the Hamiltonian for a non-commuting measurement in a 3D transmon system. The emitted photons from the cavity are collected to obtain the time-domain signals, which are then analyzed through Fourier transformation to acquire the power spectrum of the photons.
  }
\end{figure}

As the dispersive interaction yields a natural measurement in the $Z$ basis, we shift the qubit energy basis by continuously and resonantly driving the qubit at the Lamb-shifted frequency~\cite{ref:Blais2021cqe} to realize a Hamiltonian $\Omega\sigma_x/2$~\cite{ref:Blais2004cqe,ref:Krantz2019qeg}. In the doubly rotating frame of the qubit and resonator the total Hamiltonian is given by,
\begin{equation}
\label{eq:Hamiltonian}
\begin{aligned}
H_{\mathrm{R}} = \Omega \sigma_x/2 + \chi a^\dagger a \sigma_z,
\end{aligned}
\end{equation}
\xiayu{where $\Omega$ is the Rabi frequency of the qubit drive and is also the new qubit energy with eigenstates in \kater{$\sigma_x$} basis.} 

The expected energy exchange for a non-commuting measurement can be described as follows. For the two eigenstates of the qubit energy term, $\ket{+}$ and $\ket{-}$, the corresponding energy is $\Omega/2$ and $-\Omega/2$ respectively. After a $Z$~measurement that completely dephases the qubit, without knowing the measurement outcomes, the qubit would change to a fully mixed state, with zero average energy. Therefore, the qubit energy drops by $\Omega/2$ for the initial $\ket{+}$ state and increases by $\Omega/2$ for the $\ket{-}$ state. Due to energy conservation, the change of the qubit energy must be balanced by a corresponding change in the resonator photons' energy. \xiayu{As the dispersive interaction commutes with the number operator $\hat{n}=a^{\dagger}a$, the photon number in the measurement pulse remains the same. Therefore, to achieve energy conservation, the energy change of the qubit must occur as a frequency shift of the photons}, with a blue shift for qubit initially in $\ket{+}$ and a red shift for qubit initially in $\ket{-}$, as illustrated in Fig.~\ref{fig:setup}(a). \xiayu{We note that this frequency shift originates from the dephasing of the qubit during the non-commuting measurement. Thus, it is a pure quantum effect due to the measurement backaction}\kater{, which is fundamentally distinct from classical Raman scattering or wave mixing processes.} The observation and quantification of this frequency shift is the central goal of this work.

The visibility of this frequency shift depends on the energy dispersion of the pulse; for square pulses used in the experiments, if the energy dispersion is sufficiently large, i.e. a very fast measurement, the qubit state is projected in the $Z$ axis without intervening dynamics associated with the $\sigma_x$ term in the Hamiltonian, and the measurement is deemed to be ideal. Correspondingly, the energy dispersion is such that the frequency shift is not detectable. In this work, we are interested in the opposite regime, where the bandwidth of the measurement pulse is sufficiently narrow, i.e. a long pulse compared to $1/\Omega$, in order to clearly resolve the measurement energetics in terms of spectral shifts. Intriguingly, although the experimental hardware is constructed to produce a $\sigma_z$ measurement, the measurement photons acquire a spectral signature that is sensitive to the eigenstates of $\sigma_x$. The measurement is now far from ideal, but reveals the physical mechanisms behind the energy exchanges.

\section{Experimental setup} 

The experiment is realized in a 3D transmon system, which includes a superconducting transmon qubit embedded in a 3D aluminum cavity, as shown in Fig.~\ref{fig:setup}(b).
The frequency of the qubit is 5.0178~GHz in the lab frame, and an anharmonicity of $\sim$300~MHz allows us to focus exclusively on the dynamics within the qubit sub-manifold of the transmon. The qubit-cavity interaction is in the strong dispersive regime, with a dispersive shift of $\chi/2\pi = -4.0$~MHz which has a magnitude significantly larger than the cavity dissipation rate $\kappa/2\pi = 0.9$~MHz. Via this interaction, the frequency of the cavity depends on the qubit state, with $f_{g}^{\mathrm{(c)}} = 5.6959$~GHz for qubit in the ground state $\ket{g}$ and $f_{e}^{\mathrm{(c)}}=5.7039$~GHz for qubit in the excited state $\ket{e}$. The relaxation and dephasing time for the qubit are $T_1 \simeq 13.5~\mu\mathrm{s}$ and $T_2^*\simeq 2.5~\mu\mathrm{s}$, respectively. See Appendix \ref{sec:exp} for further details on the experimental setup and Hamiltonian engineering.

\begin{figure}[!t]
  \centering
  \includegraphics[width=3.2in]{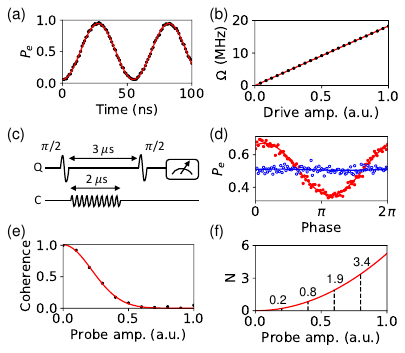}
  \caption{\label{fig:characterizatioin}(a) Rabi oscillations are used to calibrate the qubit Hamiltonian. The qubit population is measured after a qubit driving pulse with varying durations. The solid line represents a sinusoidal fit. (b) Rabi frequency as a function of the amplitude for the driving pulse. The solid line represents a linear fit. (c) The pulse sequence for the qubit (Q) and cavity (C) in a Ramsey measurement, with the final qubit measurement performed using high-power readout~\cite{ref:Reed2010hfr}. (d) The measured qubit population as a function of the second $\pi/2$ pulse's phase in the Ramsey measurement. The red solid dots (blue open dots) correspond to the results for $N=0$ ($N=1.9$) photons. The solid lines represent sinusoidal fits. The amplitude of the oscillation is proportional to the remaining qubit coherence after the measurement-induced dephasing caused by the probing pulse. (e) The remaining qubit coherence as a function of the amplitude of the probing pulse in the Ramsey measurement. Note that the coherence at $N=0$ is normalized to 1. The solid line represents a Gaussian fit with the center at zero. (f) The photon number $N$ contained in the probing pulse as a function of the pulse amplitude, calculated from the fit curve in (e). The dashed lines indicate the probe amplitude for $N=0.2$, 0.8, 1.9, and 3.4 photons.
  }
\end{figure}

Before studying the power spectra of cavity probe photons, we first calibrate the qubit energy and resonator photon number in Hamiltonian~\eqref{eq:Hamiltonian}. Square-shaped drive and probe pulses are used to respectively set the qubit energy $\Omega$ and the total emitted photon numbers $N$~\cite{ref:Linpeng2022ecm} that interact with the qubit, as illustrated in Fig.~\ref{fig:setup}(b).
For the qubit, the energy $\Omega$ is the same as the frequency of the Rabi oscillations induced by the driving pulse, which is measured at different driving amplitude as shown in Fig.~\ref{fig:characterizatioin}(a,b). For the resonator, we determine the total emitted photon number $N$ through a Ramsey experiment, as presented in Fig.~\ref{fig:characterizatioin}(c)-(f).
The axis of the second $\pi/2$ rotation in the Ramsey sequence is alternated by changing the phase of the corresponding microwave pulse, which results in a qubit population oscillation with the amplitude proportional to the remaining qubit coherence after the measurement-induced dephasing. With the probe at frequency $(f_{g}^{\mathrm{(c)}} + f_{e}^{\mathrm{(c)}})/2$, the remaining coherence is proportional to $e^{-2N}$~\cite{ref:Linpeng2022ecm, ref:Gambetta2006qpi}. We use this dependence to determine the photon number $N$ versus probe amplitude as shown in Fig.~\ref{fig:characterizatioin}(e) and (f). The probe amplitude utilized in the following paper corresponds to $N = 0.2$, 0.8, 1.9 and 3.4. Note that these photon numbers are obtained in the absence of qubit drive and are denoted as $N(\Omega=0)$ in the following.

\begin{figure*}[!t]
  \centering
  \includegraphics[width=\textwidth]{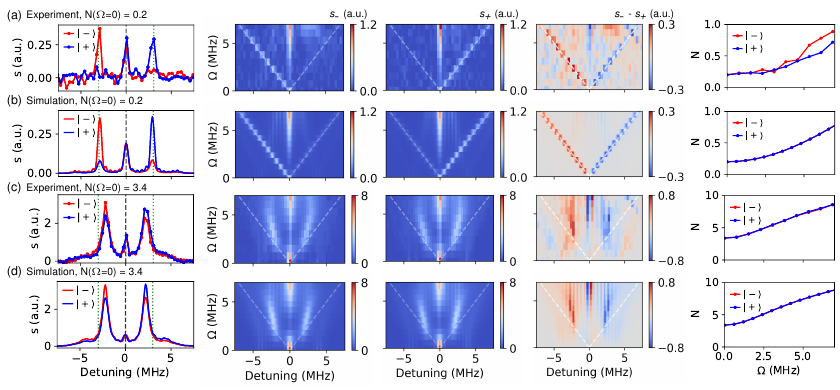}
  \caption{\label{fig:spectra} 
Power spectra of the emitted photons obtained from both experiment and simulation at photon numbers $N (\Omega=0) = 0.2$ and $N (\Omega=0) = 3.4$. The left column displays single spectra for the qubit initially in the states $\ket{+}$ and $\ket{-}$ with a qubit energy of $\Omega = 3$ MHz. The middle three columns show spectra at different $\Omega$ for the qubit initially in the $\ket{-}$ and $\ket{+}$ states, denoted as $s_{-}$ and $s_{+}$, and the difference the power spectra \kater{$s_{-}-s_{+}$}. The right column shows the photon number $N$ as a function of $\Omega$. The intensity of the simulated spectra is proportional to that of the measured spectra, but the absolute values differ as the simulated spectra is obtained from the calculated correlation function while the measured spectra is obtained from the measured voltage signal. To make the plots have similar intensity, we have multiplied the simulated spectra by a scaling factor to match the measured spectra. (a, b) Experimental and simulated results for $N(\Omega=0)=0.2$. (c, d) Experimental and simulated results for $N(\Omega=0)=3.4$.
  }
\end{figure*}

\section{Power spectrum} 

The power spectra acquired at photon numbers $N (\Omega=0) = 0.2$ and $N (\Omega=0) = 3.4$ are displayed in Fig.~\ref{fig:spectra}(a) and (c), respectively.The spectra include one central peak at the probing frequency and two side peaks located around detuning $\pm \Omega$. At small photon number, the two side peaks are precisely located at detuning $\pm \Omega$ and there is a notable distinguishability between the peak heights for qubit initially in the $\ket{+}$ and $\ket{-}$ states. For the $\ket{+}$ state, the side peak predominantly appears at $+\Omega$, indicating an average blue shift, whereas for $\ket{-}$ state, the side peak predominantly appears at $-\Omega$, indicating an average red shift. At large photon number, the side peaks and the central peak are still present, but the detuning between the two side peaks and the central peak is less than $\Omega$. This is reminiscent of the Quantum Zeno effect, whereby damping from the measurement reduces the energy scale of the qubit---effectively slowing down the quantum evolution of a system under frequent measurements ~\cite{Vijay2012,Slichter2016,Harrington2017}. Here, the applied cavity probing pulse serves as a continuous quantum measurement. 

As we have previously discussed, the expected maximum energy change of the qubit is on average $\pm \Omega/2$ for a measurement that completely projects the initial qubit state in its incompatible basis. However, we observe frequency shifts that are quantized around $-\Omega, 0, +\Omega$.
The three peaks reflect the fundamental mechanisms at play while the probe photons are scattered by the qubit-resonator system \cite{maria_and_nicolo_arxiv}. Typically, the central peak stands for trajectories where the qubit state remains unchanged, while the peak shifted by $+\Omega$ (resp. $-\Omega$) signals the qubit transition from $\ket{-}$ to $\ket{+}$ (resp. from $\ket{+}$ to $\ket{-}$). While these peaks look reminiscent to the celebrated Mollow triplet, they actually capture a very different physical situation. In particular, unlike for the Mollow triplet, there is no exchange of excitation between the qubit and the field\kater{---the photon number \xiayu{in the field} is conserved at all times.}
In a microscopic collision model analysis of the interaction of photons with the qubit we see that different photon number states acquire different frequency shifts, quantized by $\Omega$. The measurement pulse is a linear combination of different Fock states, and as such acquires a weighting of the three possible frequency shifts \cite{maria_and_nicolo_arxiv}. This is particularly apparent in the regime of strong measurement [Fig.~\ref{fig:spectra}(c)] where the contrast between the two peaks is diminished for large photon number: several photons have probed the qubit, yet the total energy change of the probe is limited to $\pm \Omega/2$.

To confirm our understanding of our measured results, we conduct a simulation of the power spectra by solving the Lindblad master equation for the system. The master equation includes the qubit $T_1$ relaxation, the qubit $T_2^*$ dephasing, and the cavity dissipation. The power spectra can be obtained by performing a Fourier transformation of the correlation function:
\begin{equation}
\label{eq:powerspec}
\begin{aligned}
s(\omega) =\frac{\kappa}{2\pi} \int^{\tau}_{0} \int^{\tau}_{0} dt_1 dt_2 e^{-i \omega (t_1-t_2)}c(t_1,t_2),
\end{aligned}
\end{equation}
where $\tau=3~\mu$s is the duration of the simulation, $c(t_1,t_2) = \langle a^\dagger(t_1) a(t_2)\rangle$ is the correlation function of the emitted photons and $a(t)$ ($a^{\dagger}(t)$) is the cavity lowering (raising) operator in the Heisenberg picture. The simulated power spectra for photon numbers $N (\Omega=0) = 0.2$ and $N (\Omega=0) = 3.4$ at different qubit energies frequencies are shown in Fig.~\ref{fig:spectra}(b) and (d), respectively. The excellent agreement between the simulations and experiments further validates the reliability of our interpretation.

As shown in Fig.~\ref{fig:spectra}, the shape of the power spectra changes with $\Omega$. In particular, with the same strength of the probe pulse, the transmitted photon number $N$ is different at different $\Omega$. The measured and simulated $N$ as a function of $\Omega$ are shown in the right column of Fig.~\ref{fig:spectra}. For both $N(\Omega=0) = 0.2$ and $N(\Omega=0) = 3.4$, $N$ increases with $\Omega$, as the central peak in the spectra becomes more predominant at larger $\Omega$, leading to higher transmission of the probe field. The measured $N$ at $\Omega' \neq 0$ is obtained through multiplying $N(\Omega=0)$ by the ratio of the integrated spectrum intensity at $\Omega'$ and the integrated spectrum intensity at $\Omega = 0$. The simulated $N$ at different $\Omega$ is directly obtained by calculating the total emitted photon number by the cavity $\int_0^{t_\mathrm{end}} \kappa n(t) dt$, where $n(t)$ is the instantaneous photon number in the cavity at time $t$. The experimental results for $N(\Omega=0) = 0.2$ case [the right-most panel of Fig.~\ref{fig:spectra}(a)] show slight difference between $\ket{+}$ and $\ket{-}$ due to relatively large noise as the signal is weak at small photon number.

\begin{figure*}[!t]
  \centering
  \includegraphics[width=6.5 in]{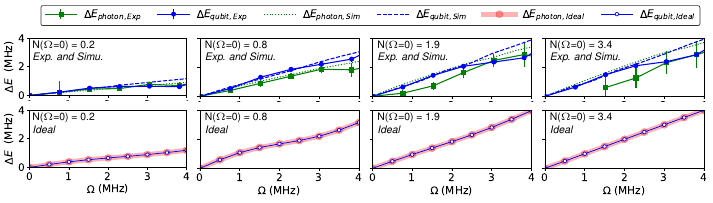}
  \caption{\label{fig:energy} 
Comparison of the qubit energy changes $\Delta E_{\text{qubit}}$ and photon energy changes $\Delta E_{\text{photon}}$ at different photon numbers: $N(\Omega=0)=0.2$, 0.8, 1.9 and 3.4. The top four panels show the results from experimental measurements and the corresponding simulations using the master equation. The error bars of $\Delta E_{\mathrm{photon, Exp}}$ are from the fit error and the error bars of $\Delta E_{\mathrm{qubit, Exp}}$ are the standard deviation from 5 repeated measurements. The bottom panels show the results from simulation in the ideal case where the qubit intrinsic decoherence is ignored and the energy changes of the reflected photons are also included \cite{supplementary}. Perfect energy conservation between the qubit and photons is observed. }
\end{figure*}

\section{Energy conservation}

Finally, we investigate the energy conservation that must be present between the energy changes in the qubit due to measurement backaction and the detected frequency shift on the probe. The qubit's energy change is determined by measuring the decrease in coherence of the qubit after the non-commuting measurement. We use a Ramsey sequence to determine the reduction in the qubit coherence, similar to Fig.~\ref{fig:characterizatioin}(c) but with qubit drive continuously on. To eliminate the influence of the intrinsic decoherence due to $T_1$ and $T_2^*$ of the qubit, we perform two Ramsey measurements, one with the cavity probe on and the other with the probe off. We estimate the remaining qubit coherence, i.e. the off-diagonal elements of the qubit density matrix by $2|\rho_{|g\rangle\langle e|}| = A_{\mathrm{on}}/A_{\mathrm{off}}$, where $A_{\mathrm{on}}$ ($A_{\mathrm{off}}$) is the measured Ramsey fringe amplitude with the probe pulse on (off). The energy change of the qubit is then calculated as
$
\Delta E_{\mathrm{qubit}} = \Omega (1 - 2|\rho_{|g\rangle\langle e|}|),
$
where the prefactor is two times $\Omega/2$ as we account for the summation of energy changes obtained with qubit initially prepared in the $\ket{+}$ and $\ket{-}$ states. Next, we obtain the energy change of the probe photons, $\Delta E_{\mathrm{photon}}$, by multiplying the photon number $N$ with the average frequency shift obtained from the power spectra with qubit initially in $\ket{+}$ and $\ket{-}$ states. 


The top four panels presented in Fig.~\ref{fig:energy} depict the measured values for four different photon numbers, along with their corresponding simulated values obtained from the solutions of the Lindblad master equation. The experimental data exhibits a good agreement with the simulation; however, neither demonstrates precise energy conservation between the qubit and photon energy changes.
This apparent violation of energy conservation comes from the fact that the joint qubit-cavity system is actually open through three main channels. First, the experiment does not have access to the reflected measurement pulse, which can carry a small energetic component. Second, the qubit is continuously and resonantly driven, which can provide a small amount of energy as soon as the qubit departs from the stationary states imposed by the drive. Finally, the qubit also undergoes thermal relaxation and dephasing making the characterization of $\Delta E_{\text{qubit}}$ inaccurate. Furthermore, a dynamical decoupling effect induced by the qubit drive effectively reduces the coherence loss of the qubit, which is not captured by the simulation using master equation. In the bottom four panels of Fig.~\ref{fig:energy}, we present the simulated results of an ideal model that includes the photon energy change in the reflected pulse and assumes no intrinsic qubit decoherence (See Appendix~\ref{sec:dispersive}), showing perfect energy conservation. \kater{Our findings affirm that the qubit energy change arising from measurement backaction during a non-commuting measurement are perfectly balanced by frequency shift of the photons utilized in the measurement.}

\section{Conclusion} We have investigated the non-resonant energetic exchange between probe and qubit arising from the incompatibility of the qubit Hamiltonian with the measurement operator. Our observations show a clear frequency shift in the power spectra of the measurement photons to ensure global energy conservation. Our results demonstrate that the energy change in the qubit is balanced by the frequency shift of the photons used in the measurement. This study elucidates the mechanisms of energy transfers between system and meter during non-commuting measurements. They provide insights for the development of future quantum measurement engines and related hardware utilizing measurement fuel.

\section*{Acknowledgements}

This work was supported by the John Templeton Foundation, Grant No. 61835, the NSF Grant No. PHY-1752844 (CAREER) and use of facilities at the Institute of Materials Science and Engineering at Washington University. MM acknowledges the support by PNRR MUR project PE0000023-NQSTI. A.A. acknowledges the National Research Foundation, Singapore and A*STAR under its CQT Bridging Grant, the Plan France 2030 through the project NISQ2LSQ ANR-22-PETQ-0006 and through the project OQuLus ANR-23-PETQ-0013, the ANR Research Collaborative Project ``QuRes" (Grant ANR-PRC-CES47-0019), and the ANR Research Collaborative Project ``Qu-DICE" (Grant ANR-PRC-CES47).








\begin{appendix}

\section{Extended details about the experimental setup} \label{sec:exp}

The Hamiltonian described in Eq.~\eqref{eq:Hamiltonian} is implemented by continuously driving the qubit, as depicted in Fig.~\ref{fig:setup}(b). \xiayu{Note that the qubit drive in the lab frame is time-dependent with a frequency of $\sim$5~GHz, thus the time-independent term $\Omega\sigma_x/2$ in Eq.~\eqref{eq:Hamiltonian} is only obtained in the rotating frame of the qubit. In the main text, unless explicitly mentioned, we study the system dynamics in the rotating frame and denote $\Omega$ as the new qubit energy, which is of MHz scale. To realize the $Z$ measurement,} a probe tone at the frequency $(f_{g}^{\mathrm{(c)}} + f_{e}^{\mathrm{(c)}})/2$, located midway between the two cavity resonances, is applied to inject cavity photons and perform the measurements. The cavity, equipped with input and output coupling ports, is weakly coupled to the environment at the input port and strongly coupled at the output port. Both the cavity probe and qubit drive are applied through the input port, with cavity photons emitted primarily through the output port. Phase-preserving amplification of the emitted photons is accomplished using a Josephson parametric amplifier (JPA), a cryogenic amplifier, and several room-temperature amplifiers. Down-conversion with a microwave $I$--$Q$ mixer produces time-domain signals for the in-phase quadrature $I(t)$ and out-of-phase quadrature $Q(t)$~\cite{ref:Krantz2019qeg}. The phasor $I(t)+iQ(t)$ is analyzed by Fourier transformation to obtain the photon power spectrum~\cite{ref:Korotkov2001cwm,ref:Korotkov2001osd}. The final spectrum is background subtracted to remove noise generated by different electronics in the collection path, using a background spectrum obtained with no cavity probing pulse. \kater{Due to the relatively high level of experimental noise, we first fit the spectrum data with three Lorentzian peaks and then extract the average frequency shift.} The experiment employs a 3~$\mu\mathrm{s}$ qubit drive pulse and a 2~$\mu\mathrm{s}$ cavity probing pulse that are simultaneously initiated, with a 1~$\mu\mathrm{s}$ gap provided to facilitate the dissipation of the cavity photons. The collected signals during the 3~$\mu\mathrm{s}$ period are used to obtain the power spectrum. 

\section{Dispersive model} \label{sec:dispersive}

We now analyze how energy is conserved in a scattering-type interaction involving two waveguides ($A$ and $B$), a cavity between these two waveguides and a qubit inside the cavity. 
We denote by $H_S$ the Hamiltonian of cavity, qubit, and their dispersive interaction. As in the main text, we consider the qubit Hamiltonian to be $H_Q = (\hbar \Omega/2)\sigma_x$. 

The experiment takes place as follows. The cavity and waveguide $B$ are initially empty while a coherent pulse travels in waveguide $A$. Since the interaction is dispersive, the qubit is isolated until the cavity fills up with photons. When the pulse arrives at the cavity, a great part is reflected and a small part enters the cavity. In the long time limit, the cavity is again empty and therefore the only energy change of $H_S$ with respect to the initial time is given by the energy change of the qubit. Indeed, at the initial and final time we consider here, the cavity contains no photons and light present in the waveguide is far away from the cavity itself. This means that, before the initial time and after the final time, the coupling term between cavity and waveguide has no effect whatsoever on the dynamics of both. This is what makes the dynamics under examination a scattering-type dynamics. It follows that the energy change of $H_S$ has to be accounted for by the change of energy in the waveguide. At initial and final times, the interaction energy is zero in the sense that its average, its variance, and every other moment of its distribution are zero.

Here we take care to analyze the role of the reflected photons in waveguide $A$, on the energy balance of the experiment. In the actual experiment, these photons are inaccessible, but we show here that their energy shifts in a different way than those emitted in waveguide $B$. The complete calculations are made in the rest of the document. The final result, i.e., the correct formula for energy conservation is:
\begin{equation}
\label{eq:EnergyConservation}
\begin{split}
-\Delta E_S = &\intmp \dd{\omega} \omega \Bigg[\left(1+\frac{\kappa_A}{\kappa_B}\right) S_B (\omega, \infty) \\
&\quad + 2 \sqrt{\kappa_A} \Re\left\{\alpha_p (\omega) \ev{c^\dagger (\omega)}_0\right\}\Bigg].
\end{split}
\end{equation}
Here, $\Delta E_S$ is the qubit energy change in the long-time limit, $\kappa_{A(B)}$ is the dissipation rate of the cavity connected to waveguides $A$ and $B$, $S_B (\omega,\infty)$ is the output spectrum in waveguide $B$, $\alpha_p (\omega)$ is the coherent input pulse in frequency space, and $\ev{c^\dg (\omega)}_0$ is the Fourier transform of $\ev{c^\dg (t)}_0$ which is the average value of the creation operator in the cavity at time $t$.


The formula of Eq.~\eqref{eq:EnergyConservation} is valid for any $H_S$. This could be a cavity with $N$ qubits inside or another cavity or something else. The only assumption is that the
interaction between cavity and waveguide is the standard one used in input-output theory and that the dynamics can be studied as a scattering-type dynamics.

\subsection{Calculations}

Our model is governed by the following Hamiltonian:
\begin{equation}
H= H_S + H_A + H_B + V_{AS} + V_{BS},
\end{equation}
where $H_S$ is the system Hamiltonian (cavity and qubit but it could be anything). For the remaining, we have (with $\hbar=1$)
\begin{gather}
H_A = \intmp \dd{\omega} \omega a^\dg (\omega) a(\omega),
\\
H_B = \intmp \dd{\omega} \omega b^\dg (\omega) b(\omega),
\\
V_{AS} = i \sqrt{\frac{\kappa_A}{2\pi}}\intmp \dd{\omega} \prt{a^\dg (\omega) c - a (\omega) c^\dg},
\\
V_{BS} = i \sqrt{\frac{\kappa_B}{2\pi}}\intmp \dd{\omega} \prt{b^\dg (\omega) c - b (\omega) c^\dg},
\end{gather}
where $c$ is the annihilation operator of the cavity in system $S$ while $a (\omega)$ and $b (\omega)$ are, respectively, the annihilation operators at frequency $\omega$ for waveguides $A$ and $B$.

By definition, the spectrum $S_A (\omega)$ is equal to $S_A (\omega) = \ev{a^\dg (\omega) a (\omega)}$. In Heisenberg picture, this means that $S_A (\omega,\tau) = \ev{a_\tau^\dg (\omega) a_\tau(\omega)}_0$, where $\tau$ denotes the time. In Heisenberg picture we get that
\begin{gather}
a_\tau (\omega) = e^{- i \omega \tau} a_0 (\omega) + \sqrt{\frac{\kappa_A}{2\pi}}\int_0^\tau e^{-i \omega (\tau - t)} c(t) \dd{t},
\\
a_\tau^\dg (\omega) = e^{+ i \omega \tau} a_0^\dg (\omega) + \sqrt{\frac{\kappa_A}{2\pi}}\int_0^\tau e^{+i \omega (\tau - t)} c^\dg (t) \dd{t},
\end{gather}
and similarly for the operators in waveguide $B$. Notice that by definition, the average energy of a waveguide at time $\tau$ is given by
\begin{align}
\ev{H_A}_\tau 
&= 
\intmp \omega \ev{a_\tau^\dg (\omega) a_\tau(\omega)}_0 \dd{\omega} \\
&=
\intmp \omega S_A (\omega,\tau) \dd{\omega}.
\end{align}

Our dynamics takes place under the assumption that there are no photons in the cavity at $t=0$ and that this cannot change until photons arrive from the waveguides. Also in the long-time limit, the cavity is again empty. It follows that the interaction is zero (in the scattering-theory sense, as explained above Eq.~\eqref{eq:EnergyConservation}) at $t=0$ and at $t=\tau$ for $\tau$ sufficiently high. So we get that, in the long-time limit:
\begin{multline}
\ev{H}_0 = \ev{H}_\tau \\
\implies
\ev{H_S}_0 + \ev{H_A}_0 + \ev{H_B}_0
=
\ev{H_S}_\tau + \ev{H_A}_\tau + \ev{H_B}_\tau
 \\ \implies
\intmp \omega \prtq{S_A (\omega,\tau) + S_B (\omega,\tau)} \dd{\omega}
=
-\Delta E_S,
\end{multline}
where $\Delta E_S = \ev{H_S}_\tau - \ev{H_S}_0$. This result is obtained under the assumption that $\ev{H_A}_0 = \ev{H_B}_0 = 0$. For $\ev{H_B}$ this happens because the waveguide is initially empty. For $\ev{H_A}$, because we put ourselves in the rotating frame such that the initial pulse has average frequency $\omega_p=0$. If this is not the case, we just have to write that
\begin{equation}
-\Delta E_S
= \intmp \omega \prtq{S_A (\omega,\tau) - S_A (\omega,0) + S_B (\omega,\tau)} \dd{\omega}.
\end{equation}

For the spectrum in waveguide $B$, we get that, at any time
\begin{equation}
 S_B (\omega,\tau) = \frac{\kappa_B}{2\pi}\int_0^\tau \int_0^\tau e^{-i (t-s)} \ev{c^\dg (t) c(s)}_0 \dd{t}\dd{s},
\end{equation}
because any occurrence of $b_0 (\omega)$ and $b^\dg_0 (\omega)$ gives zero on the vacuum state.
Regarding the waveguide $A$, the formula is a bit more complex as we get
\begin{multline}
S_A (\omega,\tau) = 
\ev{a_0^\dg (\omega) a_0 (\omega)}_0 \\ \quad \quad \quad+ \frac{\kappa_A}{2\pi}\int_0^\tau \int_0^\tau e^{-i (t-s)} \ev{c^\dg (t) c(s)}_0 \dd{t}\dd{s}
\\ \quad \quad +
\sqrt{\frac{\kappa_A}{2\pi}}  \Biggl \langle a^\dg_0 (\omega) \prt{\int_0^\tau e^{i \omega t} c(t) \dd{t}}
\\ \quad \quad \quad \quad \quad \quad \quad \quad + \prt{\int_0^\tau e^{-i \omega t} c^\dg(t) \dd{t}} a_0 (\omega) \Biggr\rangle_0.
\end{multline}
The first term is the input spectrum in waveguide $A$ and the second term is the output from the waveguide in the case when there is only emission from the cavity. We are interested in the long-time limit so we can say that $\tau \rightarrow \infty$. Since $\ev{c(t)} =0$ for $t\leq 0$ we can write
\begin{multline}
S_A (\omega,\infty) = 
\ev{a_0^\dg (\omega) a_0 (\omega)}_0 \\+
\sqrt{\kappa_A}\ev{a^\dg_0 (\omega) c(\omega) + c^\dg (\omega) a_0 (\omega)}_0 \\+
\frac{\kappa_A}{2\pi}\int \int e^{-i (t-s)} \ev{c^\dg (t) c(s)}_0 \dd{t}\dd{s},
\end{multline}
where we defined $c(\omega) = \sqrt{1/2\pi}\intmp e^{i \omega t} c(t) \dd{t}$. Exploiting the fact that $\ev{a^\dg_0 (\omega) c(\omega) + c^\dg (\omega) a_0 (\omega) }_0$\\ $= 2 \Re{\ev{c^\dg (\omega) a_0 (\omega)}_0}$ and that the input field in waveguide $A$ is a coherent field we obtain
\begin{multline}
S_A (\omega,\infty) = \ev{a_0^\dg (\omega) a_0 (\omega)}_0 \\+ 2 \sqrt{\kappa_A} \Re{\alpha_p (\omega) \ev{c^\dg (\omega)}_0} + \frac{\kappa_A}{\kappa_B} S_B (\omega,\infty),
\end{multline}
where $\ev{c^\dg (\omega)}_0 = \sqrt{1/2\pi} \intmp e^{i \omega t} \ev{c^\dg(t)}_0 \dd{t}$. Notice that $S_A (\omega, 0) = \ev{a_0^\dg (\omega) a_0 (\omega)}_0$. Finally, the energy conservation reads
\begin{multline}
-\Delta E_S = \intmp \dd{\omega} \omega \Biggl[\prt{1+\frac{\kappa_A}{\kappa_B}} S_B (\omega, \infty) \\+ 2 \sqrt{\kappa_A} \Re{\alpha_p (\omega) \ev{c^\dg (\omega)}_0}\Biggr].
\end{multline}

\end{appendix}

\end{document}